\documentclass[aps,pra,twocolumn,groupeaddress,showpacs,floatfix]{revtex4}

\usepackage{amsmath}
\usepackage{amsfonts}
\usepackage{amssymb}
\usepackage{bm}
\usepackage{bbm}
\usepackage{graphicx}
\usepackage{epsfig}
\usepackage{latexsym}
\usepackage{nicefrac}
\usepackage{color}
\usepackage{dcolumn}

\usepackage[NoRemarks]{optional}

\def\corresponds{{\lower.1.377ex\hbox{=}}{\rm\kern-.75em^\triangle}}
\def\succsim{\succ\kern-.9em_\sim\kern.3em}
\def\precsim{\prec\kern-1em_\sim\kern.3em}
\def\slantfrac#1#2{\kern1em^{#1}\kern-.3em/\kern-.1em_{#2}}
\def\lfrac#1#2{{}^{#1\!}\kern-.0em/_{#2}}

\def\buildrel#1\under#2{\mathrel{\mathop{\kern0pt #2}\limits_{#1}}}

\definecolor{light}{gray}{0.90}
\definecolor{darker}{gray}{0.50}
\definecolor{dark}{gray}{0.30}

\def\ii{{\mathrm{i}}}

\def\tfrac#1#2{ {\textstyle{\frac{#1}{#2}} } }

\newcolumntype{.}{D{x}{}{7}}

\newcommand{\addrMST}{
Department of Physics, Missouri University of Science and Technology,
Rolla, Missouri, MO65409-0640, USA}

\newcommand{\addrMTADE}{
MTA--DE Particle Physics Research Group,
P.O.Box 51, H--4001 Debrecen, Hungary}

\begin{document}

\sloppy

\title{Light Sea Fermions in Electron--Proton and Muon--Proton Interactions}

\author{U. D. Jentschura}
\affiliation{\addrMST}
\affiliation{\addrMTADE}

\begin{abstract} 
The proton radius conundrum 
[R.~Pohl {\em et al.}, Nature {\bf 466}, 213 (2010) and A.~Antognini
{\em et al.} Science {\bf 339}, 417 (2013)]
highlights the need to revisit any conceivable sources of 
electron-muon nonuniversality in lepton-proton interactions 
within the Standard Model.  Superficially, a number of perturbative
processes could appear to lead to such a nonunversality.  One of these is a
coupling of the scattered electron into an electronic as opposed to a muonic
vacuum polarization loop in the photon exchange of two valence quarks,
which is present only for electron projectiles as opposed to muon 
projectiles.  However, we can show
that this effect actually is part of the radiative correction to the proton's
polarizability contribution to the Lamb shift, equivalent to a radiative
correction to double scattering. We conclude that 
any conceivable genuine nonuniversality must be connected
with a nonperturbative feature of the proton's structure, e.g., with the
possible presence of light sea fermions
as constituent components of the proton.
If we assume an average of roughly $0.7 \times 10^{-7}$ light sea positrons per valence
quark, then we can show that 
virtual electron-positron annihiliation processes lead to an extra
term in the electron-proton versus muon-proton interaction, which has the right
sign and magnitude to explain the proton radius discrepancy.  
\end{abstract}


\maketitle

%
%
\section{Introduction}
\label{sec1}

The electromagnetic aspects of the proton and neutron structure 
are somewhat elusive. It is well known that 
the mass difference of proton and neutron is responsible for 
the stability of the universe
(the hydrogen atom would otherwise be unstable 
against beta decay into an electron-positron pair, a neutrino and 
a neutron). There have been attempts to explain the 
mass difference on the basis of the electromagnetic 
interaction among the 
constituent quarks~\cite{ChEtAl1974,ChTh1976,GoHaJa1983,MoSa1985}.
A priori, one would think that the 
electrostatic interaction among the 
constituent quarks leads to an inversion of the 
mass hierarchy of proton versus neutron.
Namely, the Coulomb interaction among 
valence quarks actually lowers the energy of the neutron as compared 
to the proton, as a naive counting argument shows.
A hadron with valence quarks $uud$ has 
interquark electromagnetic interactions proportional to 
the fractional charge numbers,
$\tfrac23 \times \left( -\tfrac13 \right) +
\tfrac23 \times \left( -\tfrac13 \right) +
\tfrac23 \times \tfrac23 = 0$ 
whereas for the neutron with valence quarks $udd$,
we have
$\tfrac23 \times \left( -\tfrac13 \right) +
\tfrac23 \times \left( -\tfrac13 \right) +
\left( -\tfrac13 \right) \times \left( -\tfrac13 \right) = -\tfrac13$.
The latter expression, being negative, would suggest that the 
neutron is lighter than the proton if the mass difference
were of electromagnetic origin and due to Coulomb exchange.

However, the radiative correction is not constrained to have any particular
sign, and warrants further investigation especially because the electromagnetic
wave functions of the valence quarks bound in an MIT bag
model~\cite{ChEtAl1974} have a rather peculiar structure~\cite{MoSa1985} and
might give rise to significant radiative effects.  The conclusion reached in
Refs.~\cite{ChEtAl1974,ChTh1976,GoHaJa1983,MoSa1985} is that the
electromagnetic self-energy of the quarks remains positive for all masses
considered.  Thus, the quantum electrodynamic (QED) radiative energy shift
cannot explain the mass difference of proton and neutron, where a negative
self-energy would otherwise be required in view of the larger fractional charge
of the up as compared to the down quarks.

Still, the investigations~\cite{ChEtAl1974,ChTh1976,GoHaJa1983,MoSa1985}
as well as the proton radius conundrum~\cite{PoEtAl2010,AnEtAl2013} 
highlight the need for a closer look at the internal structure of the proton if
one is interested in its own ``internal'' electromagnetic interactions, as well
as the interactions of the proton with the ``outside world''.  
If the interaction of the bound or scattered lepton with the proton is nonuniversal,
then it is conceivable that the proton radius depends on the 
projectile particle.
However, one can show that a number of 
perturbative higher-order effects which could appear
to lead to such a nonuniversality of electron-proton versus muon-proton 
interactions are in fact absorbed into correction terms of known 
physical origin.

Let us consider electromagnetic interactions among the constituent particles of
the proton, for example,
a higher-order effect generated by a coupling of the scattered projectile
(electron or muon) into a vacuum-polarization loop which in turn is inserted
into a photon exchanged between two valence quarks. We 
here show that, because
Feynman propagators take care of all possible time orderings of virtual
particle creation and annihilation processes, this effect actually 
constitutes a radiative correction to double scattering, and 
is absorbed, in Lamb shift calculations, into the radiative correction
to the proton's polarizability contribution to the Lamb shift.
A quantitative parametric 
estimate for the order-of-magnitude of the effect is provided.

The second process is more
speculative and conjectures the presence of light sea fermions as a
nonperturbative physical property of the hadron, an admixture to
the genuine particle content of the proton.
We here show that the conceivable presence of these
fermions would give rise to a Dirac-$\delta$ potential, in view of a virtual
annihilation channel, with the right sign to explain the muonic hydrogen
puzzle~\cite{PoEtAl2010,AnEtAl2013}. These two mechanism are described in the
following two Secs.~\ref{sec2} and~\ref{sec3}, respectively.
Units with $\hbar = c = \epsilon_0$ are used throughout this paper
unless stated otherwise.

\begin{figure}[t!]
\begin{center}
\begin{minipage}{0.99\linewidth}
\begin{center}
\includegraphics[width=1.0\linewidth]{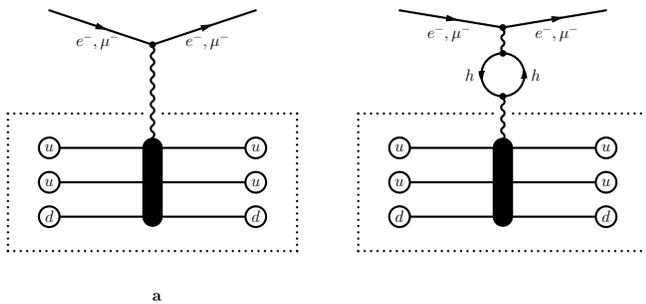}
\end{center}
\caption{\label{fig1} Diagram~(a) is the standard
scattering process involving an incoming electron or 
muon, without radiative corrections. The hadronic 
vacuum-polarization contribution from diagram~(b) 
is taken into account consistently in Lamb shift calculations 
and subtracted from scattering data as a radiative correction.
The photon is emitted ``collectively'' by the quarks
inside the proton. The proton as a compount particle 
is encircled by dotted lines.}
\end{minipage}
\end{center}
\end{figure}

\begin{figure}[t!]
\begin{center}
\begin{minipage}{0.99\linewidth}
\begin{center}
\includegraphics[width=0.7\linewidth]{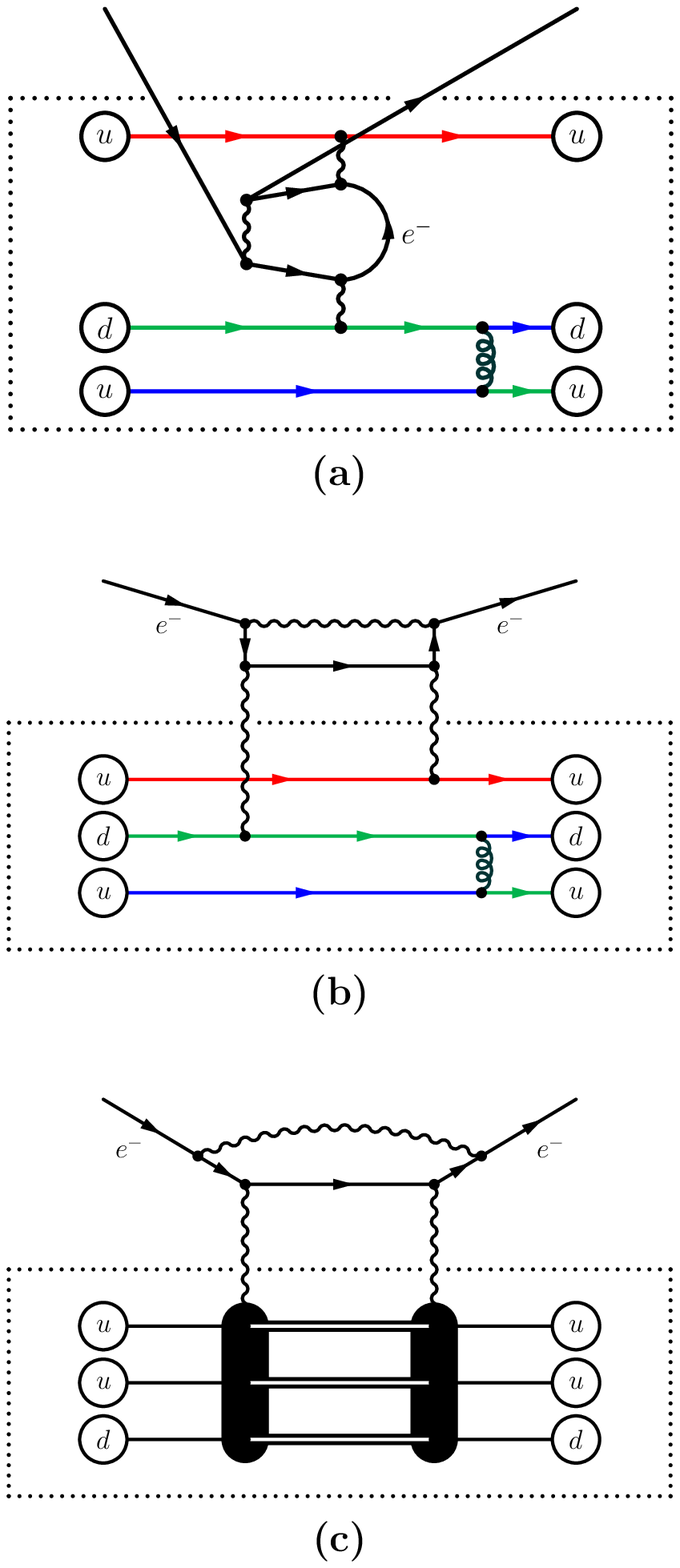}
\end{center}
\caption{\label{fig2} (Color online.) Diagram~(a)
describes the coupling of an incoming electron into the 
vacuum-polarization loop of an electromagnetic interquark interaction 
inside the proton. Feynman propagators describe all possible 
time orderings of the particle creation and annihilation 
processes and diagram~(b) thus describes the same effect as~(a).
The gluon interaction in diagram (b) is representative.
The inelastic contribution to diagram~(b), with an ``excited'' quark in the 
virtual state, is identified in diagram~(c) as a radiative 
correction to the proton's polarizability contribution
to the Lamb shift.}
\end{minipage}
\end{center}
\end{figure}

%
%
\section{Lepton--Proton Scattering and Internal Structure of the Proton}
\label{sec2}

Let us first recall the relevant conventions.  The leading-order process for
the scattering of leptons (e.g., electrons or muons) off of a proton is
depicted in Fig.~\ref{fig1}(a).  A virtual photon is emitted
collectively by the proton, describing the electromagnetic interaction of 
lepton and hadron.  Any insertions of virtual particles into the exchange
photon are absorbed into the $F_1$ and $F_2$ form
factors (or Sachs $G_E$ and $G_M$ form factors) of the proton, while the proton
radius is defined as the slope of the Sachs $G_E$ form factor, with all those
terms [and radiative corrections, see Fig.~\ref{fig1}(b)]
subtracted from $G_E$. These would otherwise be ascribed to a point proton with
the properties of a structureless spin-$1/2$ Dirac particle. 

Let us briefly review the status (see also Ref.~\cite{Je2011radii}). 
The proton interaction 
vertex is changed, in view of the nontrivial form factors, as
\begin{equation}
\gamma^\mu \to
\gamma^\mu \, F_1(q^2) + 
\frac{\ii \sigma^{\mu\nu} q_\nu}{2 m_p} \, F_2(q^2) \,,
\end{equation}
where $F_1$ and $F_2$ are the Dirac and Pauli form factors 
of the proton, respectively.
The electric and magnetic $G_E$ and $G_M$ Sachs form factors are defined 
in terms of the $F_1$ and $F_2$ as follows,
\begin{subequations}
\begin{align}
G_E(q^2) =& \; F_1(q^2) + \frac{q^2}{4 (m_p c)^2} \, F_2(q^2) \,,
\\[0.007ex]
G_M(q^2) =& \; F_1(q^2) + F_2(q^2) \,, \qquad
F_2(0) = \varkappa_p \,.
\end{align}
\end{subequations}
One canonically separates the Sachs $G_E$ form factor into a 
QED contribution $G_E^{\rm QED}(q^2)$, which 
captures all aspects of the point-particle QED 
nature of the proton,
and a part $\overline G_E(q^2)$ 
which is due to the proton's internal structure~\cite{Je2011radii},
\begin{equation}
G_E(q^2) = \overline G_E(q^2) + G_E^{\rm QED}(q^2) \,.
\end{equation}
The definition of the proton charge radius then reads
as
\begin{equation}
\langle r^2 \rangle_p =
6 \hbar^2 \left.  \frac{\partial \overline G_E(q^2)}{\partial q^2} 
\right|_{q^2 = 0} \,,
\end{equation}
i.e.~it measures the internal structure of the 
proton, after all QED contributions (``radiative corrections'') 
have been subtracted (and that includes the infrared divergent 
slope of the QED one-loop contribution to the $F_1$ form factor).
By definition, the subtraction of the 
QED contribution $G_E^{\rm QED}(q^2)$ also includes 
all virtual loop insertions into the exchange photon line 
that would otherwise affect the proton-lepton interaction for a 
point proton. One of these is the hadronic vacuum-polarization loop 
in Fig.~\ref{fig1}(b).

The proton radius is measured in the low-energy 
region, where one can use a dipole fit to 
$ \overline G_E(q^2)$ to good approximation.
Let us briefly recall the fundamental differences
of low-energy elastic scattering, which mainly 
determines the proton's size, and high-energy deep
inelastic scattering  (DIS), 
which is relevant for momentum transfers $q^2 \gg m_p^2$.
For an incoming lepton four-momentum $\ell_1$ 
and an incoming proton momentum $p$
(outgoing lepton momentum $\ell_2$ and exchange photon 
four-momentum $q$), the Bjorken scaling law~\cite{Bj1969}
is as follows.
After the subtraction of radiative 
correction, one writes the deep inelastic cross section as
\begin{align}
\sigma_{\rm DIS} =&\; \sigma_0 \, 
\left[ \frac{G^2_E(Q^2, \nu) + \tau \, G^2_M(Q^2, \nu)}{1 + \tau} 
\right.
\nonumber\\[0.007ex]
& \; \left. + 2 \tau \, G^2_M(Q^2,\nu) \, \tan^2\left( \frac{\theta}{2} \right) \right]
\nonumber\\[0.007ex]
\equiv & \;  \sigma_0 \, \left[ W_2(Q^2,\nu) + 2 \, W_1(Q^2, \nu)
\, \tan^2\left( \frac{\theta}{2} \right) \right] \,,
\end{align}
where $\sigma_0$ is the Mott scattering cross section,
$\nu = q \cdot p/m_p$ is the energy loss of 
the lepton, $Q^2 \equiv - q^2$, $\tau = Q^2 (4 m_p^2)^{-1}$,
and $\theta$ is the scattering angle  of the lepton,
i.e., the angle subtended by the spatial components of $\ell_1$ and $\ell_2$.
The form factors $G_E(Q^2, \nu)$ and $G_M(Q^2, \nu)$
describe inelastic scattering (with energy loss), and the elastic
counterparts are recovered in the limit $\nu \to 0$.
Bjorken~\cite{Bj1969} observed that, if the scattering in the high-energy region 
were to come from point-like constituents inside the proton, 
then the structure functions $W_1$ and $W_2$ should be 
consistent with scattering from asymptotically free 
constituents (``partons'' or ``quarks''),
\begin{subequations}
\begin{align}
\lim_{\mbox{\scriptsize $\begin{array}{c}
Q^2 \to \infty\\ Q^2/\nu \; \mbox{const.}
\end{array}$}} \nu \, W_2(Q^2, \nu) = & \; m_p \, F_2(x) \,,
\\[0.007ex]
\lim_{\mbox{\scriptsize $\begin{array}{c}
Q^2 \to \infty\\ Q^2/\nu \; \mbox{const.}
\end{array}$}} W_1(Q^2, \nu) = & \; F_1(x) \,,
\quad 
x \equiv \frac{Q^2}{2 m_p \nu} \,,
\end{align}
\end{subequations}
where the $F_1$ and $F_2$ are now structure
functions instead of form factors; their argument is 
the Bjorken $x$ variable.
The Bjorken scaling was confirmed by the famous 
SLAC--MIT experiments~\cite{BlEtAl1969,Ta1991,Ke1991,Fr1991}.
However, in dealing with low-energy scattering processes
and contributions to the Lamb shift, the proton's form factor 
can be approximated very well using a dipole fit 
[see, e.g.,~the discussion surrounding Eq.~(74) of Ref.~\cite{Pa1996mu}].

Let us consider the diagram in Fig.~\ref{fig2}(a), which could superficially be
assumed to induce a nonuniversality of the electron-proton versus muon-proton
interaction, on the level of higher-order corrections.  Namely, the coupling of
the projectile electron into the {\em electronic} vacuum-polarization loop of an
electromagnetic interquark interaction is available only for an incoming
electron (as opposed to an incoming muon).  The Feynman propagators for the
fermions and the leptons in Fig.~\ref{fig2}(a) contain all possible time
orderings, including scattering ``backward in time'' which leads to the
vacuum-polarization loop.  The diagram in Fig.~\ref{fig2}(b) thus describes the
same physical process as Fig.~\ref{fig2}(a).
Furthermore, it is necessary to remember what the ``scattering
off of a definite quark inside the proton'' [see Fig.~\ref{fig2}(a)]
physically means in the characteristic momentum range of an 
electron or muon bound to the proton. It implies that the 
proton's internal state changes in between the two interactions 
of the virtual electron with the virtual photons emitted 
by the proton [see Fig.~\ref{fig2}(c)].
Thus, the process in Fig.~\ref{fig2}(a) can finally be identified 
as a radiative correction to proton's polarizability 
contribution to the Lamb shift, as depicted in Fig.~\ref{fig2}(c).

The contribution of double-scattering processes is canonically subtracted in
the analysis of scattering experiments. In the context of bound states, the
leading contribution from two-photon exchange (without radiative corrections)
gives rise to the so-called third Zemach moment term which is proportional to a
convoluted charge distribution of the proton~\cite{Fr1979,Je2011radii,MoGrSa2013}.
The elastic
correction to the third Zemach moment due to the proton structure can be taken
into account by inserting proton form factors into the two-photon exchange
forward scattering amplitude [Eqs.~(70)--(75) of Ref.~\cite{Pa1996mu}], and the
inelastic correction to the third Zemach moment (due to an excited state of the
proton in between the photon exchanges, also known as the proton polarizability
correction) is numerically too small to explain the proton 
radius puzzle~\cite{Pa1996mu,Je2011aop1,Je2011aop2,Bo2012,AnEtAl2013aop}.

Finally, let us provide a parametric estimate for the 
contribution of the diagram in Fig.~\ref{fig2}(c),
based on the analogy with the two-Coulomb-vertex 
correction to the self-energy, as given by the calculation 
reported in Ref.~\cite{BaBeFe1953}.
The induced effective potential
for the diagram in Fig.~\ref{fig2}(c),
by scaling arguments, can be estimated to be proportional to 
\begin{equation}
H_{\rm vp} \propto \frac{[\alpha_{\rm QED}(m_{\rm eff}^2/m_e^2)]^3}{m_{\rm eff}^2} \,
\delta^{3}(r) \,,
\end{equation}
Here, $\alpha_{\rm QED}$ is the running 
QED coupling which is approximately equal to $1/137.036$ at 
zero momentum transfer,
$\delta^3(r)$ is the three-dimensional Dirac-$\delta$ function,
and $m_{\rm eff}$ is an effective mass or momentum scale entering the 
loop in Fig.~\ref{fig2}(c). 
The latter can be estimated as follows.
Let $\lambda \sim r_p$ be a characteristic de Broglie wavelength 
of the quarks inside the nucleus. 
Then, the associated momentum scale is 
$p \sim h/r_p$ where $h$ is Planck's unit of action
and the corresponding energy scale is obtained as $E \sim p \, c
\sim 1.32 \, m_p$, which in turn is commensurate
with the excitation energy of the proton into
its first resonance, the $\Delta$ resonance at $1232\,{\rm MeV}$.
It is easy to check, based on the running of the QED coupling,
that the effective coupling at the scale of the proton's momentum
differs from the value of $\alpha_{\rm QED}$ at zero momentum
transfer by less than 5\%.
The leading finite-size Hamiltonian is given as follows,
\begin{equation}
\label{Hsize}
H_{\rm fs} = \frac{2 \pi \alpha_{\rm QED}}{3 m_e^2} \,
\left[ m_e^2 \, \left< r_p^2 \right> \right] \, \delta^3(r) \,.
\end{equation}
The ratio is given as
\begin{equation}
R \propto 
\frac{\langle H_{\rm vp} \rangle}{ \langle H_{\rm fs} \rangle}
\sim 
\alpha^2_{\rm QED} \frac{m_e^2}{m_{\rm eff}^2} \,
\frac{1}{m_e^2 \, \left< r_p^2 \right>} \sim 2.2 \times 10^{-6} \,,
\end{equation}
where we take into account that 
$m_e^2 \, \left< r_p^2 \right> \sim (1/386)^2$,
and $m_e/m_{\rm eff} \sim m_e/m_p \sim 5.4 \times 10^{-4}$.
The ratio $R$ is too small to make a significant contribution 
to a solution of the proton radius puzzle.
It is interesting to note that the simple-minded
parametric estimate described above, with one 
radiative factor $\alpha_{\rm QED}$ from the self-energy 
loop excluded, gives the right order-of-magnitude for the 
leading proton polarizability contribution 
evaluated in Ref.~\cite{MoGrSa2013}.

%
%
\section{Light Sea Fermions}
\label{sec3}

\begin{figure}[t!]
\begin{center}
\begin{minipage}{0.99\linewidth}
\begin{center}
\includegraphics[width=0.91\linewidth]{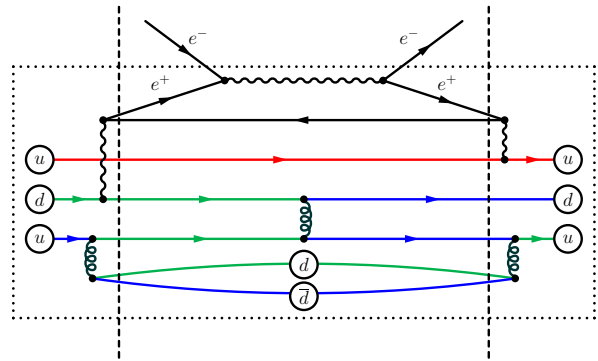}
\end{center}
\caption{\label{fig3} (Color online.) Typical Feynman diagram illustrating the virtual
annihilation of a bound electron with a ``light sea lepton'' (positron) inside
the proton.  The up ($u$) and down ($d$) quarks, which carry non-integer charge
numbers, interact electromagnetically.
The dashed lines mark the
formation of the
asymptotic state of the proton in the distant past or future, with
its valence and sea quark contents, and with a light sea lepton
that annihilates with the bound electron.
The given Feynman diagram is not
included if the proton is treated perturbatively as a spin-$1/2$ particle with
charge $e$. An exemplary quantum chromodynamic
(QCD) interaction via a blue-antigreen gluon also is
indicated in the figure. }
\end{minipage}
\end{center}
\end{figure}

Let us consider the possible presence of light sea fermions
as nonperturbative contributions to the proton's structure,
inspired by a possible importance of virtual electron-positron
pairs in the lepton nonuniversality in interactions with protons.
We consider a thought experiment:
If we switched off the electroweak
interactions of quarks inside the proton, the resulting ``proton''
would of course be neutral but otherwise rather comparable in its mass
and in its nuclear properties to a real proton with some
nonperturbative quantum chromodynamic (QCD)
``wave function.'' Now, if we include back the
electroweak interactions of quarks, virtual photons and
electron-positron pairs would backreact on the previous
``wave function'' leading to a reshaping, and the
actual ``proton wave function'' which now additionally contains photons
and the electron-positron pairs.
Due to highly nonlinear nonperturbative nature of QCD, 
this reshaping can be much larger than the electromagnetic perturbation itself, and
therefore there is a room for the conceivable presence
of electron-positron pairs inside the proton, which cannot be accounted for 
by perturbative QED considerations alone 
(see Fig.~\ref{fig3}). This density (probability) 
of electron-positron pairs,
because of the  inherently nonperturbative nature of QCD, is difficult 
if not impossible to quantitatively estimate,
but its presence is not excluded by any known experiments.
In fact, a significant photon content of the proton is well confirmed
in the so-called Deep Inelastic Compton 
Scattering
(DICS, see Refs.~\cite{CoKe1992,Bl1993,MuPi2004prd,MuPi2004epjc,dRVo1999,LeSCWe2003}
and Fig.~\ref{fig3}).

If the proton contains these
electron-positron pairs, which are not accounted for in any 
perturbative higher-order QED term, 
then the interaction between the proton and electron
is given by both photon exchange and annihilation diagrams.
In natural units,
the photon annihilation diagram in the case
of positronium leads to the effective interaction~\cite{BeLiPi1982vol4}
\begin{equation}
\delta H = \frac{\pi \alpha_{\rm QED}}{2 m_e^2} \, 
\left( 3 + \vec \sigma_+ \cdot \vec \sigma_-
\right)  \, \delta^3(r) \,.
\end{equation}
This Hamiltonian gives
a nonvanishing interaction of the bound electron and the light sea positron
if their spins add up to one.  Assuming that the
electron-positron pairs within the proton are not polarized, we can replace
$\vec \sigma_+ \cdot \vec \sigma_- \to 0$ after averaging over the polarizations
of the sea leptons. For atomic (electronic) hydrogen, the additional
interaction of the electron with the proton due to the annihilation channel 
therefore is of the form
\begin{equation}
\label{Hann}
H_{\rm ann} = \epsilon_p \, \frac{3 \pi \alpha_{\rm QED}}{2 m_e^2} \, \delta^3(r) \,,
\end{equation}
where $\epsilon_p$ measures the amount of electron-positron pairs
within the proton. For muonic hydrogen, the effect is expected
to vanish because the dominant contribution to the sea leptons
comes from the lightest leptons, namely, electron-positron pairs
and thus the annihilation channel is not available.
By comparison, the finite nuclear size effect is
given by Eq.~\eqref{Hsize}.
For an $S$ state, the ratio of the corresponding energy shifts is
\begin{equation}
\label{epsp}
\frac{ \left< H_{\rm ann} \right> }{ \left< H_{\rm fs} \right> }
= \frac94 \; \frac{\epsilon_p}{ m_e^2 \, \left< r_p^2 \right> }
\mathop{=}^{!} \frac{0.88^2 - 0.84^2}{0.88^2}
= 0.089 \,.
\end{equation}
The equality marked with the exclamation mark has to hold
if we are to explain the discrepancy of the electronic and
muonic hydrogen values of the proton charge radius, which
are roughly $0.88 \, {\rm fm}$ and $0.84 \, {\rm fm}$,
respectively~\cite{MoTaNe2008,PoEtAl2010,AnEtAl2013,Je2011aop1}.
The parameter $\epsilon_p$ thus can be as low as
\begin{equation}
\epsilon_p = 2.1 \times 10^{-7} \,,
\end{equation}
and still explain the effect the different proton radii  obtained
from electronic and muonic hydrogen. Per valence quark,
one thus has a fraction of 
$\epsilon_p/3 = 0.7 \times 10^{-7}$ sea fermion pairs.
The interaction due to
the annihilation channel has the right sign: it enhances the nuclear size
effect for electronic as opposed to muonic hydrogen and
thus makes the proton appear larger for electronic systems.

%
%
\section{Conclusions}
\label{sec4}

Let us include some historical remarks.
In the 1970s, transition frequencies in muonic
transitions were found to be in disagreement with theory~\cite{DiEtAl1971}.
After a sign error in the calculation
of the two-loop vacuum polarization correction~\cite{Fr1969} was
eliminated~\cite{Bl1972,BrMo1978}
and a standard $\gamma$-ray spectrometer used in the experiments was
recalibrated~\cite{DeKeSaHe1980},
other experiments later found
agreement of theory and experiment in muonic systems (e.g.,
Refs.~\cite{TaEtAl1978,BeEtAl1986}). 
Nuclear radii of some carbon, nitrogen and oxygen isotopes~\cite{DuEtAl1974}
were determined by analyzing
muonic transitions, and the resulting radii were found to 
be in agreement with electron
scattering radii to better than $5\,\%$.  Later,
the radius of ${}^{12}{\rm C}$ was updated~in
Refs.~\cite{RuEtAl1984,OfEtAl1991}, finally 
``converging'' to a value of $r_C = 2.478(9) \,
{\rm fm}$, in good agreement with the value from muonic x-ray studies.
Muonic atom and ion spectroscopy is meanwhile regarded 
an established tool for the determination of nuclear radii~\cite{AnEtAl2009}.

However, the light sea fermions discussed in Sec.~\ref{sec3} are distributed
only inside the nucleons as opposed to the entire nucleus which is held
together by meson exchange, because the local electromagnetic 
(EM) field is strongest inside the
protons and neutrons. The size of the proton could be determined by the light
sea fermions, among other things, but the size of a composed large nucleus is
determined by the nuclear meson-mediated force.  Expressed differently, the
muonic hydrogen experiments probe {\em one and only one} nucleon, whereas other
experiments involving, say, a muon bound to a ${}^{12}{\rm C}$ nucleus, probe
the charge radius of an ensemble of nucleons, which is 
mainly determined by the
arrangement of the nucleons inside the ${}^{12}{\rm C}$~nucleus.  Thus, the
effect proposed in Sec.~\ref{sec3} should become smaller for larger nuclei.
A rough estimate would entail the 
observation that the carbon nucleus is three times bigger than the proton.
So, naively and classically, three nucleons fit into 
the diameter of the ${}^{12}{\rm C}$~nucleus. If every one
of these has its effecetive diameter reduced by 5\%, then the 
overall radius is
reduced by only 1.7\%.  While the tables of Ref.~\cite{OfEtAl1991}
suggest agreement to (slightly) better than 1\% 
for independent experimental determinations of the 
${}^{12}{\rm C}$ charge radius, it is noteworthy that this
agreement has been achieved only after earlier discrepancies 
had been resolved.

Very interestingly, a possible electron-muon nonuniversality has been seen in a
scattering experiment~\cite{CaEtAl1969} some fourty years ago, comparing the
scattering of electrons versus muons off of protons, and was cautiously
ascribed by the authors of Ref.~\cite{CaEtAl1969} to an incorrect normalization
of the scattering data.  The observed 8\% difference in the cross sections
observed in Ref.~\cite{CaEtAl1969} translates into a 4\% difference in the
proton radius, with the same sign and magnitude as that observed in muonic
spectroscopy experiments~\cite{PoEtAl2010,AnEtAl2013}.  If one ignores the
possiblility of an incorrect normalization of the data in
Ref.~\cite{CaEtAl1969}, then the proton, ``seen'' with the ``eyes'' of a muon,
appears to be 4\%$\div$5\% smaller as compared to its ``appearance'' when
``seen'' through the ``eyes'' of an electron~\cite{CaEtAl1969,PoEtAl2010,AnEtAl2013}.
The experiment~\cite{CaEtAl1969}
urgently needs to be confronted with an independent investigation. 

According to Refs.~\cite{Je2011aop2,JaRo2010} and other theoretical works which
came to the same conclusion, it is hard to imagine any perturbative process
(direct exchange of a virtual ``subversive'' particle, or insertion of a
``subversive'' particle into the exchange photon line) which could explain the
muonic hydrogen discrepancy without seriously questioning the validity of other
measurements and corresponding theory, such as the muon $g$ factor measurement.
Furthermore, any other perturbative insertions into the photon-proton vertex,
conceivably involving internal constituents of the proton, are absorbed in the
definition of the proton radius and thus could not explain the discrepancy (see
Sec.~\ref{sec2}).  Without questioning the validity of the Maxwell equations or
quantum electrodynamics (QED), and without introducing any additional
virtual particles, it is perhaps permissible to speculate
that a nonperturbative mechanism such as the one
proposed in Sec.~\ref{sec3} might be a feasible candidate in the case of
further experimental confirmations of the proton radius
discrepancy~\cite{CaEtAl1969,PoEtAl2010,AnEtAl2013}
between electronic as opposed to muonic bound systems.

\vfill

{\bf Note added.} The experimental observation of slightly smaller 
cross sections in muon-proton versus electron-proton scattering
has been made indepdendently in Refs.~\cite{CaEtAl1969,BrEtAl1972}.
One may consult the first row of graphs in Fig.~15 of 
Ref.~\cite{BrEtAl1972} (which pertain to elastic as opposed
to inelastic cross sections), the general remarks made 
in Sec.~VII of Ref.~\cite{BrEtAl1972},  and the discussion surrounding
Eq.~(48) of Ref.~\cite{BrEtAl1972}, which is consistent with 
an 8\,\% difference in the electron versus muon cross sections.

\section*{Acknowledgments}

The author acknowledges insightful discussions with Professor K.~Pachucki.
Helpful conversations with Professor K.~Meisser are also gratefully
acknowledged. The author wishes to acknowledge support from the National
Science Fundation (grant PHY--1068547) and from the National Institute of
Standards and Technology (precision measurement grant).

\end{document}